\documentclass[fleqn,usenatbib]{mnras}

\usepackage{newtxtext,newtxmath}

\usepackage[T1]{fontenc}
\usepackage{ae,aecompl}

\usepackage{graphicx}	
\usepackage{amsmath}	
\usepackage{amssymb}	

\title[Diffuse radio emission in SPT-CL J2031-4037]{Diffuse radio emission in the galaxy cluster SPT-CL J2031-4037: a steep spectrum intermediate radio halo?}

\author[R. Raja et al.]{
Ramij Raja,$^{1}$\thanks{E-mail: phd1601121008@iiti.ac.in}
Majidul Rahaman,$^{1}$
Abhirup Datta,$^{1}$
Jack O. Burns,$^{2}$
\newauthor
H. T. Intema,$^{3,4}$
R. J. van Weeren,$^{3}$
Eric J. Hallman,$^{2}$
David Rapetti,$^{2,5,7}$
\newauthor
and Surajit Paul$^{6}$
\\
$^{1}$Discipline of Astronomy, Astrophysics and Space Engineering, Indian Institute of Technology Indore, Simrol, 453552, India\\
$^{2}$Center for Astrophysics \& Space Astronomy, Department of Astrophysical \& Planetary Sciences, 389 UCB,\\ University of Colorado,
Boulder, CO 80309, USA\\
$^{3}$Leiden Observatory, Leiden University, PO Box 9513, 2300 RA Leiden, The Netherlands \\
$^{4}$International Centre for Radio Astronomy Research, Curtin University, GPO Box U1987, Perth, WA 6845, Australia\\
$^{5}$NASA Ames Research Center, Moffett Field, CA 94035, USA\\
$^{6}$Department of Physics, Savitribai Phule Pune University, Pune 411007, India\\
$^{7}$Universities Space Research Association, Mountain View, CA 94043, USA
}

\date{Accepted XXX. Received YYY; in original form ZZZ}

\pubyear{2015}

\begin{document}
\label{firstpage}
\pagerange{\pageref{firstpage}--\pageref{lastpage}}
\maketitle

\begin{abstract}
The advent of sensitive low frequency radio observations has revealed a number of diffuse radio objects with peculiar properties that are challenging our understanding about the physics of the intracluster medium.
Here, we report the discovery of a steep spectrum radio halo surrounding the central Brightest Cluster Galaxy (BCG) in the galaxy cluster SPT-CL J2031-4037.
This cluster is morphologically disturbed yet has a weak cool core, an example of cool core/non-cool core transition system, which harbours a radio halo of $\sim 0.7$ Mpc in size.
The halo emission detected at 1.7 GHz is less extended compared to that in the 325 MHz observation, and the spectral index of the part of the halo visible at 325 MHz to 1.7 GHz frequencies was found to be $-1.35 \pm 0.07$.
Also, $P_{1.4\ \mathrm{GHz}}$ was found to be $0.77 \times 10^{24}$ W Hz$^{-1}$ which falls in the region where radio mini-halos, halo upper limits and ultra-steep spectrum (USS) halos are found in the $P_{1.4\ \mathrm{GHz}} - L_\mathrm{X}$ plane.
Additionally, simulations presented in the paper provide support to the scenario of the steep spectrum.
The diffuse radio emission found in this cluster may be a steep spectrum \lq\lq intermediate\rq\rq\ or \lq\lq hybrid\rq\rq\ radio halo which is transitioning into a mini-halo.
\end{abstract}




\section{Introduction}
In the hierarchical structure formation scenario, smaller units like galaxies or groups merge to form large scale structures in the cross-sections of cosmic web filaments.
Major merger events are the most energetic phenomena since the Big Bang, releasing about $10^{64}$ ergs of energy within a Gyr timescale.
This massive amount of energy dissipates into the intracluster medium (ICM) primarily via shocks and large scale turbulent motion (e.g.~\citealt{Sarazin2002,Paul2011}). 

Traditionally, cluster-wide diffuse radio synchrotron emission has been divided into three categories: Giant Radio halos (GRHs), Relics, and mini-halos (MHs) \citep{Feretti1996}.
GRHs are centrally located, unpolarized Mpc scale emission sources found only in merging clusters, whereas MHs are smaller versions ($\sim 100-500$ kpc) but found only in \lq\lq relaxed\rq\rq\ cool core clusters.
However, recent discoveries of diffuse radio objects, especially at low frequencies, with properties in between GHRs and MHs have made these classifications more complicated (see \citealt{vanWeeren2019}).

The origin of GRH is merger driven turbulence in the ICM \citep{Brunetti2001, Petrosian2001}, in this scenario, less energetic events (e.g., minor mergers or mergers in smaller systems) are predicted to generate steep-spectrum halos \citep{Cassano2010}.
On the other hand, MHs are formed due to the turbulence caused by the later stage of a minor merger event i.e., sloshing core \citep{Mazzotta2008}.
An \lq\lq intermediate\rq\rq\ state between these was also proposed by \citet{Brunetti2014} where a radio halo transitioning into a mini-halo or vice-versa and recent discoveries of $\sim$ Mpc scale radio halos in cool core and semi-relaxed clusters seems to favour this scenario (e.g., \citealt{Bonafede2014,Sommer2017, Savini2018,Kale2019}).

In this letter, we report the discovery of diffuse radio emission in the SPT-CL J2031-4037 (hereafter SPT2031) cluster. 
This cluster was first detected in the REFLEX (ROSAT-ESO Flux Limited X-ray) Galaxy Cluster survey as reported by \citet{Bohringer2004}.
Subsequent detections of SPT2031 via the Sunyaev-Zel'dovich effect were reported by \citet{Plagge2010,Williamson2011,Bleem2015} and \citet{Planck2016}.
SPT2031 is a massive cluster with $M_{500}$ = $(9.83 \pm 1.5) \times 10^{14}$ $M_{\odot}$ \citep{Bleem2015} and X-ray luminousity of $L_{[0.1-2.4\ \mathrm{keV}]} = 10.4 \times 10^{44}$ erg s$^{-1}$~\citep{Piffaretti2011} situated at redshift $z = 0.3416$~\citep{Bohringer2004}.
The global cluster properties are presented in Table \ref{tab:global_prop}.

In this letter, we assumed a $\Lambda$CDM cosmology with $H_0 = 70$ km s$^{-1}$ Mpc$^{-1}$, $\Omega_{m} = 0.3$ and $\Omega_\Lambda = 0.7$. At the cluster redshift $z = 0.3416$, $1\arcsec$ corresponds to 4.862 kpc.

\begin{table}
	\caption{Global cluster and halo properties}
	\label{tab:global_prop}
	\scalebox{1.0}{
	\begin{tabular}{lc} 
		\hline
        RA & 20h31m51.5s \\
        DEC & -40d37m14s\\
        $z$ & 0.3416\\
        $R_{500}$ [Mpc] & 1.342\\ 
        $M_{500}$ [$10^{14}\ M_{\odot}$] & $9.83 \pm 1.5$\\ 
        $L_{[0.1-2.4\ \mathrm{keV}]}$ [$10^{44}$ erg s$^{-1}$] & 10.4\\
        $T_{\mathrm{central}}$ [keV] & $12.2 \pm 2.4$\\
        \hline
        $S^{\mathrm{halo}}_{325\ \mathrm{MHz}}$ [mJy] & $16.93 \pm 1.76$\\
        $S^{\mathrm{halo}}_{1.7\mathrm{GHz}}$ [mJy] & $1.4 \pm 0.18$\\
        $P^{\mathrm{halo}}_{1.4\mathrm{GHz}}$ [$10^{24}$ W Hz$^{-1}$] & $0.77 \pm 0.08$\\
        $\alpha^{1679}_{325}$ [halo] & $-1.35 \pm 0.07$\\
		\hline
	\end{tabular}}
\\References: \citet{Birzan2017, Bleem2015, McDonald2013, Piffaretti2011}.
\end{table}

\begin{figure*}
	\begin{tabular}{cc}
	\includegraphics[scale=0.32]{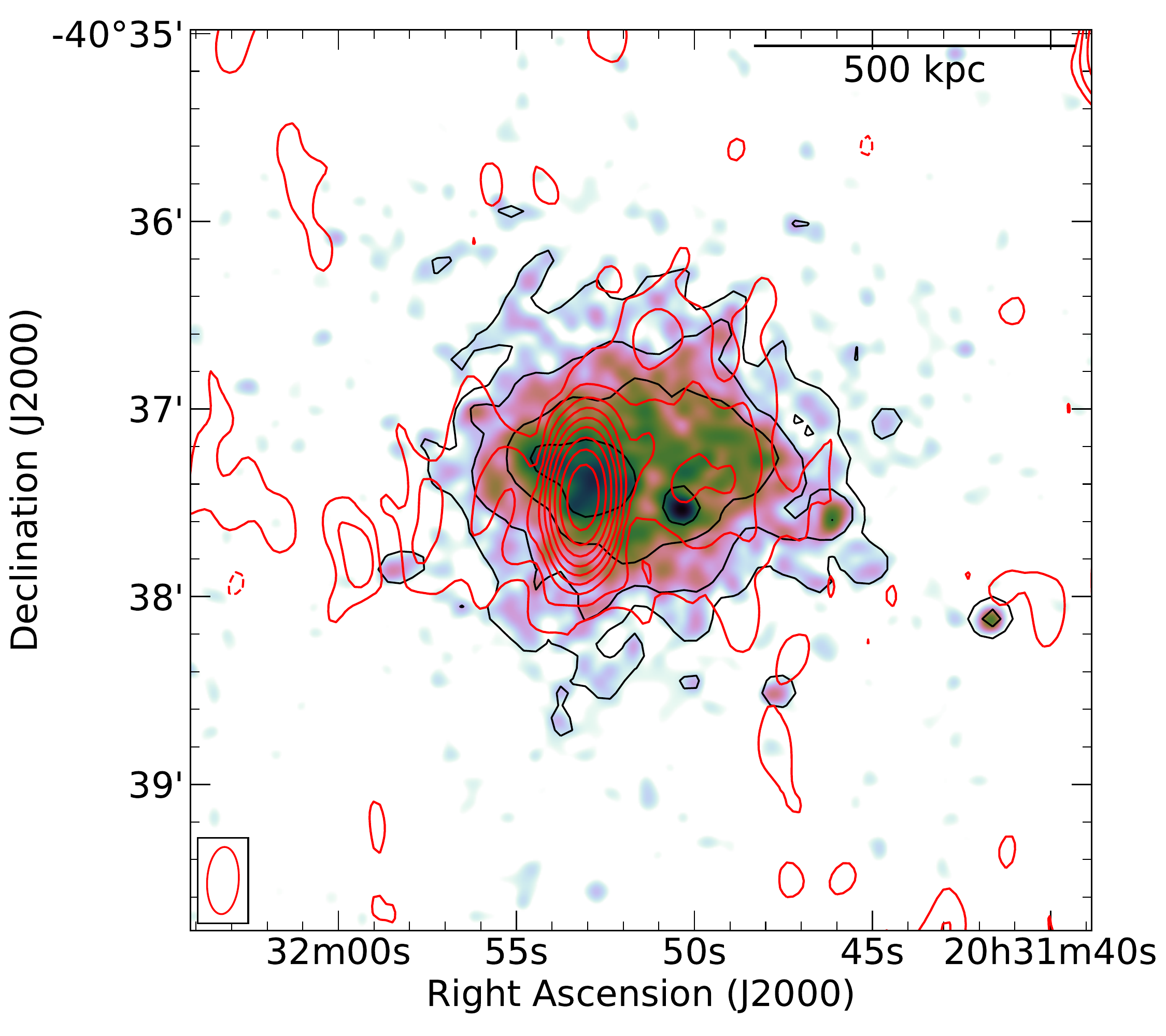} &
	\includegraphics[scale=0.32]{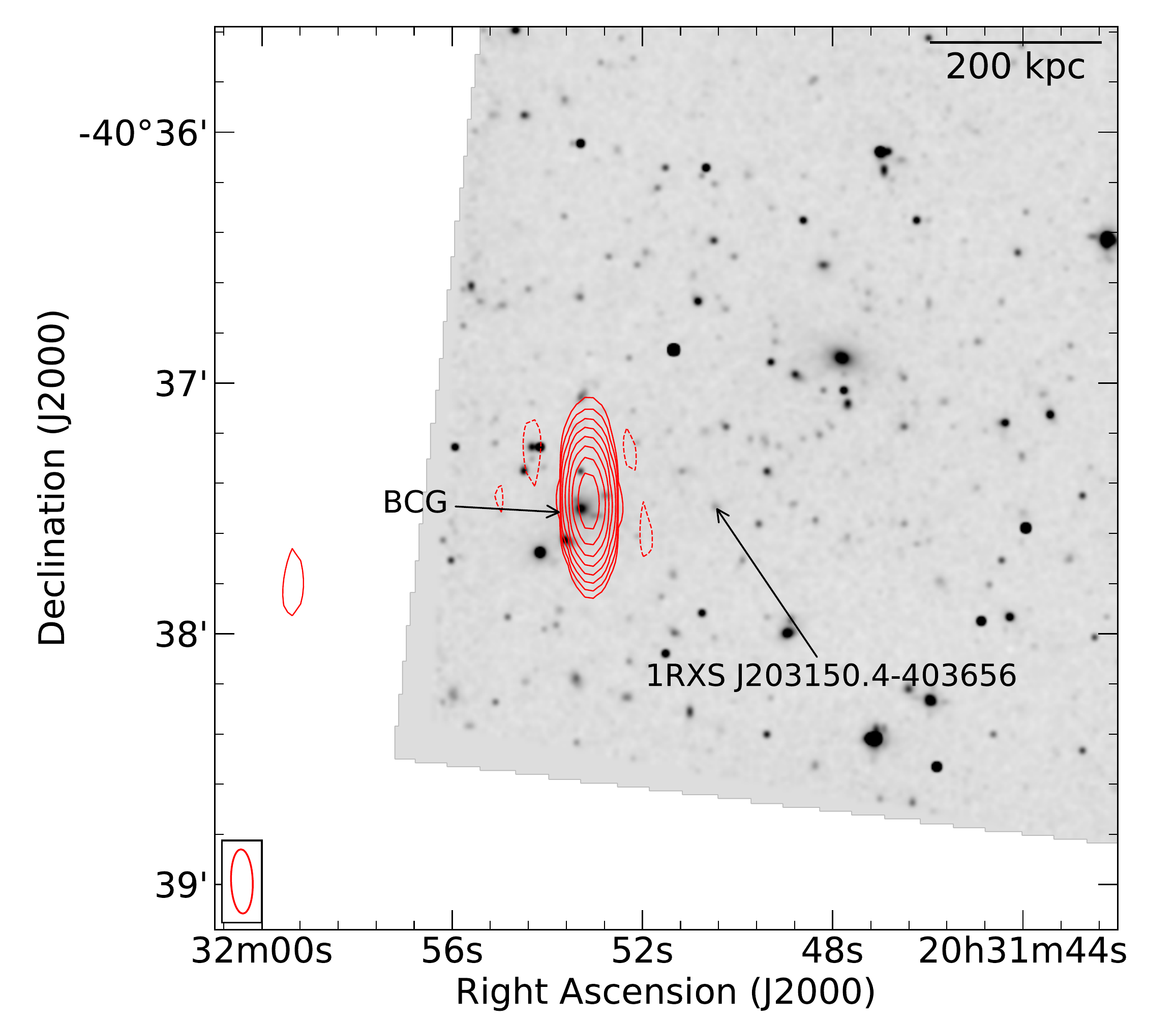}
	\end{tabular}
    \caption{{\it Left:} GMRT 325 MHz radio contours (red) of the SPT2031 cluster overlaid onto the {\it Chandra} X-ray image. The restoring beam of this image is $21.5\arcsec \times 10.1\arcsec$, PA $-3.5^{\circ}$ indicated at the bottom left corner. The red contours are drawn at levels $[-1, 1, 2, 4, 8,...] \times 3\sigma_{\mathrm{rms}}$ with $\sigma_{\mathrm{rms}}$ = 60 $\mu$Jy beam$^{-1}$. Negative contours are indicated with dotted lines. The black X-ray contours are spaced by a factor of 2. The bright X-ray source west of the BCG, 1RXS J203150.4-403656, which also has optical identification in the HST image, is situated at $z = 0.351$ and therefore not part of this cluster. {\it Right:} HST optical image overlaid with the 325 MHz radio contours excluding baselines shorter than $5$k$\lambda$. The contours are drawn at the same levels as previously but with $\sigma_{\mathrm{rms}}$ = 120 $\mu$Jy beam$^{-1}$. 
    }
    \label{fig:P_diff}
\end{figure*}

\section{Radio Data Analysis} \label{Obs}

\subsection{GMRT Observations} \label{Obs_GMRT}
GMRT observations of SPT2031 were performed at 325 MHz as part of a larger survey of a sample of SPT clusters.
It was observed on 31 May, 21 and 23 August 2014 for a total of about 6.15 hrs of on-source observing time with 32 MHz bandwidth divided into 256 channels.

The {\tiny SPAM}
pipeline was used for data reduction.
It is a python based {\tiny AIPS}
(Astronomical Image Processing System) extension to reduce low frequency interferometric data, developed by \citet{Intema2009,Intema2017}.
The data reduction process starts with initial flagging, bandpass and gain calibration loops.
The flux density of the calibrator 3C48 was set according to the \citet{Scaife2012} scale.
Then, several rounds of self-calibration were performed on the calibrated data along with RFI (Radio Frequency Interference) flagging.
Widefield facet imaging was performed to correct for the non-coplaner array.
Finally, direction dependent calibration was performed for the bright sources present in the field (for details see \citealt{Intema2009,Intema2017}).
The final calibrated visibility data were used for further imaging with {\tiny CASA}
 (Common Astronomy Software Applications; \citealt{McMullin2007}).
\subsection{VLA Observation} \label{Obs_VLA}
An L-band ($1-2$ GHz) observation of SPT2031 was made with EVLA CnB configuration on 9 Jan 2015 for a total of $\sim$1.5 hrs of on-source observing time with 1 GHz of total bandwidth. 

{\tiny CASA} was used for VLA data reduction and imaging, and a brief description of the procedure is given below. 
First, we applied the RFI mitigation software AOFlagger developed by \citet{Offringa2010, Offringa2012} to get rid of the data contaminated by RFI. 
Next, we run the VLA calibration pipeline\footnote{\url{https://science.nrao.edu/facilities/vla/data-processing/pipeline}} 
on the data, which performed basic calibration and flagging using {\tiny CASA}. A detailed description of the pipeline steps can be found in the link below.
3C48 was used as gain and bandpass calibrator, and the flux scale was set according to \citet{Perley2013}.
Few spectral windows were affected by RFI very badly and were dropped in the following steps. 
Visual inspection was done to remove any remaining bad data.
Finally, a few rounds of phase-only self-calibration was performed to remove residual phase errors.
\section{Results}
\subsection{325 MHz data}

The {\it Chandra} X-ray image in Figure \ref{fig:P_diff} shows the disturbed nature of the cluster stretching in the east-west direction. 
The GMRT 325 MHz radio contours reveal the diffuse radio emission surrounding the central radio galaxy, typical of radio mini-halos, but the emission has similar east-west stretch as in the X-ray, roughly tracing the possible merger axis like giant radio halos.
The position of this radio galaxy coincides with the X-ray luminous core, and the Hubble Space Telescope (HST) optical image clearly shows the BCG as the optical counterpart.
The BCG emission was also detected at the 150 MHz TIFR GMRT Sky Survey (TGSS; \citealt{Intema2017}) with a flux density of $S_{150} = 232.6 \pm 24.3$ mJy and
at the 843 MHz Sydney University Molonglo Sky Survey \citep{Bock1999,Mauch2003} with a flux density of $S_{843} = 42.1 \pm 1.8$ mJy.

The 325 MHz image of the SPT2031 cluster is presented in Figure \ref{fig:P_diff} with red contours.
This is a full resolution image with Briggs robust parameter \citep{Briggs1995} set to 0.5.
The total extent of the diffuse emission is $2.7\arcmin \times 2.1\arcmin$ or $0.79 \times 0.62$ Mpc.
To estimate the integrated flux density of the diffuse radio emission at 325 MHz, we used the following approach.
First, we made an image excluding baselines shorter than 5 k$\lambda$ (corresponding to a physical scale of about 
200 kpc) and using robust = -1 to get rid of the extended emission.
After subtracting the modelled BCG flux density from the uv-data, we re-imaged the diffuse radio emission. We estimated the total radio emission enclosed within the $3\sigma_{\mathrm{rms}}$ contour to be $S_{\mathrm{halo},325\ \mathrm{MHz}} = 16.93 \pm 1.76$ mJy.
The uncertainty in the flux density measurement was estimated based on the map noise and assuming an absolute flux calibration uncertainty of $10\%$ (e.g., \citealt{Cassano2007}).


\subsection{$1-2$ GHz data}
\begin{figure}
    \centering
    \resizebox{0.85\hsize}{!} {\includegraphics[width=\columnwidth]{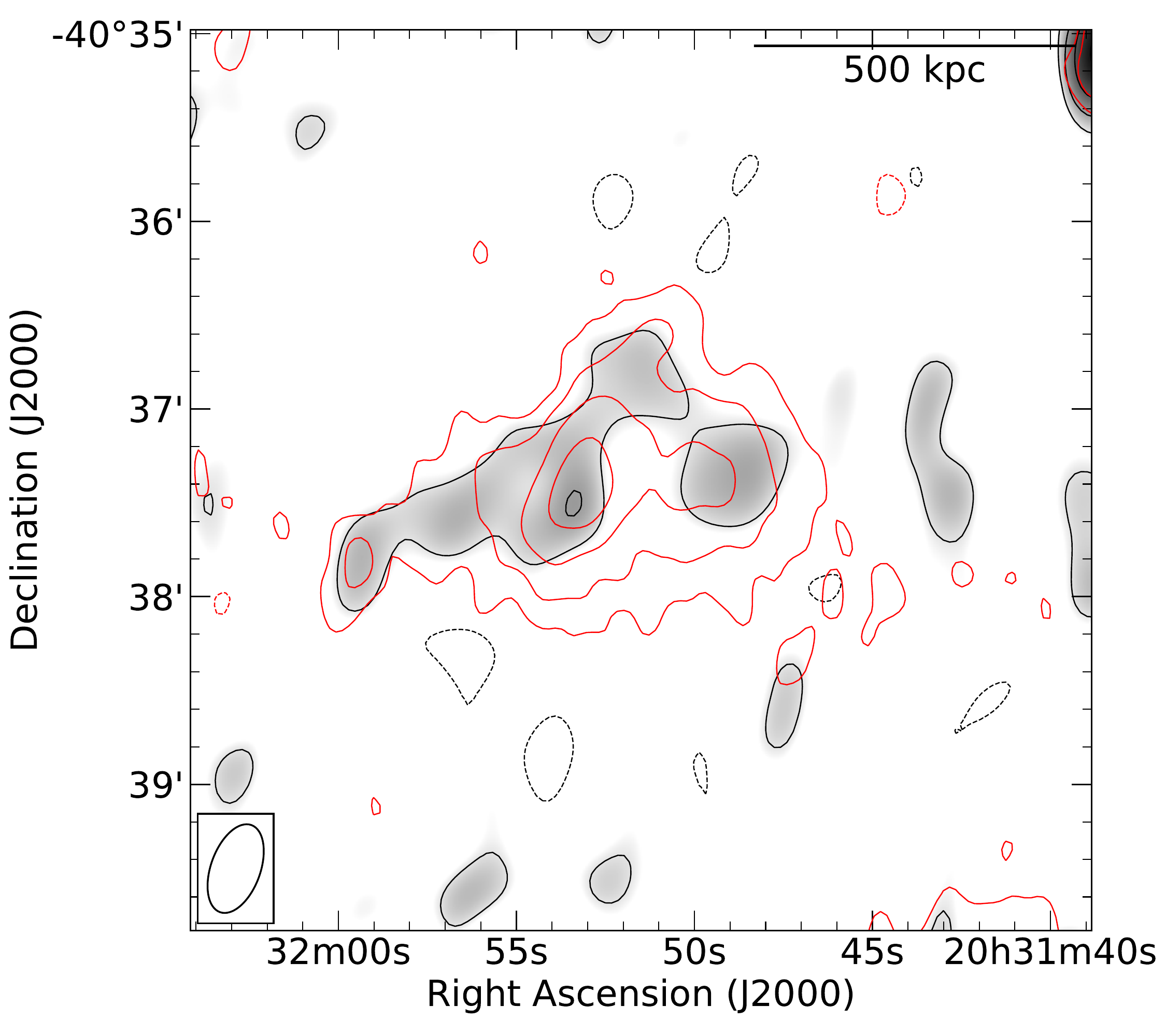}}
    \caption{VLA 1.7 GHz BCG subtracted low resolution image of the SPT2031 cluster overlaid with BCG subtracted 1.7 GHz contours (black) and 325 MHz contours (red). 
    The restoring beam of these images is $29.6\arcsec \times 15.8\arcsec$, PA $-19.84^{\circ}$ indicated at the bottom left corner. Contours are drawn at levels $[-1, 1, 2, 4, 8,...] \times 3\sigma_{\mathrm{rms}}$ with $\sigma_{\mathrm{rms}}$ = 35 $\mu$Jy beam$^{-1}$ (black) and $\sigma_{\mathrm{rms}}$ = 100 $\mu$Jy beam$^{-1}$ (red). Negative contours are indicated with dotted lines.}
    \label{fig:L_diff}
\end{figure}

In Figure \ref{fig:L_diff}, we present the 1.7 GHz observation of the SPT2031 cluster.
MS-MFS \citep{Rau2011} imaging was performed using {\tiny CASA} task {\it CLEAN} with 2 Taylor terms for spectral modeling.
This is a low resolution image, made using 5 k$\lambda$ uv-taper and setting robust = 0.5 to emphasize the diffuse emission sensitivity.
The 1.7 GHz image also shows a similar east-west stretch of the diffuse emission as in the 325 MHz observation.
The linear size of this emission at 1.7 GHz is $0.72 \times 0.47$ Mpc.
Following the similar procedure as in 325 MHz case we estimated the flux density of the diffuse emission to be $S_{\mathrm{halo},1.7\mathrm{GHz}} = 1.4 \pm 0.18$ mJy.


\subsection{Spectral index estimation} \label{subsec:spix_estimate}
Since the diffuse emission at 1.7 GHz is not as extended as in 325 MHz, we have calculated the spectral index taking precisely the same region where radio emission is present in both frequencies.
The 325 MHz image was smoothed with the restoring beam of the 1.7 GHz image to match the resolution of both images.
The spectral index between frequencies 325 MHz and 1.7 GHz came out to be $\alpha^{1679}_{325} = -1.35 \pm 0.07$ (where $S_{\nu} = \nu^{\alpha}$) 
and the corresponding k-corrected 1.4 GHz radio power is $P_{1.4\mathrm{GHz}} = (0.77 \pm 0.08) \times 10^{24}$ W Hz$^{-1}$.
Some amount of large scale diffuse emission observed at 325 MHz is not detected at 1.7 GHz, and only an upper limit to the spectral index can be derived.
To derived an upper limit to the spectral index, we followed a method similar to as described in \citet{Kale2015}. We injected a $2\arcmin \times 1.5\arcmin$ disk into the uv-data with the {\tiny CASA} task \textit{FT} and \textit{UVSUB}. The injected flux densities were calculated from the less bright part of the 325 MHz diffuse emission that is not detected at 1.7 GHz, scaled with varying spectral index values. Starting from spectral index -1.35, we lowered it up to -1.52 where the injected diffuse source can no longer be considered detected. So, the upper limit to the spectral index of steep spectrum diffuse emission is $\alpha^{1679}_{325} \leq -1.52$.


\section{Cluster Dynamical state}

To understand the origin of cluster-wide synchrotron emission, it is necessary to know the dynamical state of the cluster.
For the morphological classification of SPT2031, we checked different classification parameters widely used in the literature.
Broadly speaking, cluster dynamical state is classified as to whether the cluster hosts a cool core or not and how disturbed is the cluster.
One of the most popular classification parameters to test for the presence of a cool core is the surface brightness \lq\lq concentration\rq\rq\ parameter $c_{\mathrm{SB}}$ introduced by \citet{Santos2008}.
\citet{McDonald2013} reported $c_{\mathrm{SB}} = 0.05$ for SPT2031.
The threshold for non-cool core (NCC) is $c_{\mathrm{SB}} < 0.075$ \citep{Santos2008} which categorizes this cluster as NCC.
\citet{Hudson2010} found a tri-modal distribution of clusters for both central specific \lq\lq entropy\rq\rq\ $K_0$ and \lq\lq cooling time\rq\rq\ $t_{\mathrm{cool},0}$.
In this distribution, CC/NCC partition in cooling time is at $\sim7.7\ h^{-\frac{1}{2}}$ Gyr and strong cool core (SCC)/weak cool core (WCC) partition is at $\sim1\ h^{-\frac{1}{2}}$ Gyr.
In the case of entropy, SCC clusters have low central entropy ($\lesssim30\ h^{-\frac{1}{3}}$ keV cm$^2$), WCC clusters around $50\ h^{-\frac{1}{3}}$ keV cm$^2$, and NCC clusters have $>110\ h^{-\frac{1}{3}}$ keV cm$^2$.
$K_0$ and $t_{\mathrm{cool},0}$ for SPT2031 are 189.8 keV cm$^2$ and 3.43 Gyr \citep{McDonald2013}, which classifies this cluster as NCC and WCC respectively.
Apart from that, \citet{Morandi2015} classified CC/NCC systems depending on the density $E^{-2}_\mathrm{z} n_{\mathrm{e},0}$ at $0.03 R_{500}$ and classified SPT2031 as a CC system.

Parameters used to estimate the disturbance in a cluster are
e.g., \lq\lq centroid shifts\rq\rq\ ($w$; \citealt{Mohr1993}), photon \lq\lq asymmetry\rq\rq\ ($A_{\mathrm{phot}}$; \citealt{Nurgaliev2013}).
\citet{Nurgaliev2017} reported $w = 0.017$ for this cluster which classifies it as a disturbed cluster ($> 0.01$) although it does lie close to the threshold boundary.
Even though, $A_{\mathrm{phot}} = 0.25$~\citep{Nurgaliev2017} falls in the moderate asymmetry range ($0.15-0.6$) corresponding to a moderately disturbed cluster, it lies closer to the relaxed threshold (0.15) than the disturbed threshold (0.6).

The combination of all these morphological parameters suggests that SPT2031 is somewhere in between CC and NCC system.
A summary of these morphology classifications is presented in Table \ref{tab:morphology}.
Furthermore, \citet{Hudson2010} suggested $t_{\mathrm{cool},0}$ as the best proxy for low redshift cool core (CC) systems and \citet{Nurgaliev2013} suggested $A_{\mathrm{phot}}$ being preferable over $w$ to measure cluster disturbance for its stability with respect to the observational signal-to-noise.
Accordingly, the SPT2031 cluster can be classified as a moderately disturbed, weak cool core (WCC) cluster.

\begin{table}

\caption{Dynamical state}
\label{tab:morphology}
\scalebox{1.0}{

\begin{tabular}{lccc} 
\hline
Parameter & Value & Morphology & Ref.\\
\hline
$c_{\mathrm{SB}}$ & $0.05^{+0.00}_{-0.02}$ & NCC & 1\\
$t_{\mathrm{cool},0}$ [Gyr] & $3.43^{+0.75}_{-0.72}$ & WCC & 1\\
$K_0$ [keV cm$^2$] & $189.8^{+39.9}_{-38.9}$ & NCC & 1\\
$E^{-2}_\mathrm{z} n_{\mathrm{e},0}$ at $0.03 R_{500}$ & $-$ & CC & 2\\
$w$ & $0.017^{+0.001}_{-0.002}$ & Disturbed & 3\\
$A_{\mathrm{phot}}$ & $0.25 \pm 0.04$ & Moderate & 3\\
 & & disturbed & \\

\hline
\end{tabular}}
\\References: (1) \citet{McDonald2013}, (2) \citet{Morandi2015}, (3) \citet{Nurgaliev2017}. 
NCC, WCC and CC represent Non Cool Core, Weak Cool Core and Cool Core respectively.
\end{table}

\section{Discussion}
\label{sec:origin}
\citet{Burns2008} showed that it is hard to destroy the cool core of a cluster with a merger event at a later stage of its evolution.
So, even if a merger happened in the past in the SPT2031 cluster, it was unable to destroy the cool core completely as evident from the X-ray brightness contours (Figure \ref{fig:P_diff}) but disturbed the ICM nonetheless. 

The merger induced turbulence can re-accelerate {\it in-situ} relativistic electrons necessary to produce the extended emission found in the SPT2031 cluster.
A piece of compelling evidence for that is the similar east-west stretch of the radio emission in the direction of the merger axis (Figure \ref{fig:P_diff}).
The radio emission at 325 MHz is brighter on the west side of the BCG where the ICM is more disturbed compared to the east, which further strengthens this correlation.
The extent of the halo emission at 325 MHz is about $\sim 600 - 800$ kpc, covering most of the cluster region visible in X-ray.
However, much of the halo emission is not visible at 1.7 GHz. 

\begin{figure}
    \centering
    \resizebox{0.85\hsize}{!} {\includegraphics[width=\columnwidth]{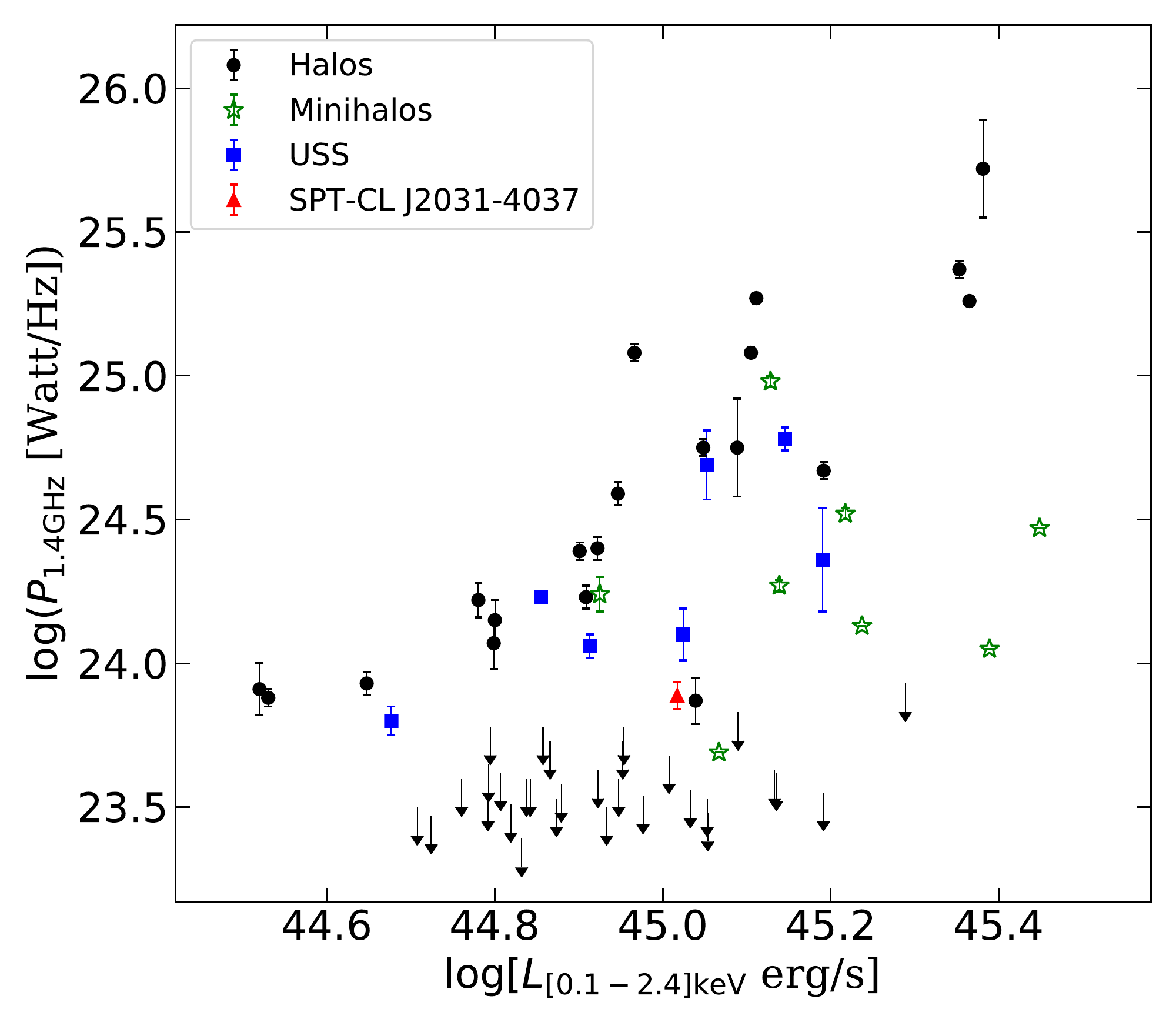}}
    \caption{The distribution of halos, mini-halos, USS halos and halo upper limits (black arrows) in the $L_\mathrm{X} - P_{1.4\mathrm{GHz}}$ plane (\citealt{Cassano2013,Kale2015}) is presented here. The SPT2031 cluster is indicated with a red triangle.}
    \label{fig:corr}
\end{figure}

According to Figure 2 of \citet{Nurgaliev2017}, $A_{\mathrm{phot}} = 0.25$ corresponds to a post-merger time of about $\sim$2 Gyr.
The spectral index of the halo emission visible in both frequencies i.e., along the merger axis is about $\alpha^{1679}_{325} = -1.35 \pm 0.07$ whereas the upper limit to the off-axis region is $\alpha^{1679}_{325} \leq -1.52$.
The derived radio power is below the expected level from the GRH $L_\mathrm{X} - P_{1.4\mathrm{GHz}}$ correlation and falls in a region between the ultra-steep spectrum (USS) halos, mini-halos and radio halo upper limits (Fig. \ref{fig:corr}).
So, the scenario could be that the observed halo emission is caused by a past less energetic merger event and after $\sim$Gyr has passed, the radio emission at a higher frequency is visible only along the merger axis where energy injection was highest.
Since the cool core is not destroyed and the radio BCG is coincident with cool core, the late stage of this merger event can trigger sloshing motion in the ICM, resulting in a mini-halo emission.
Consequently, the diffuse emission in SPT2031 of $\sim$0.7 Mpc size surrounding the BCG in a WCC system can be classified as a steep spectrum \lq\lq intermediate\rq\rq\ radio halo \citep{vanWeeren2019} which is in transition into a possible future mini-halo.
Also, the SPT2031 cluster is massive enough to host a radio mini-halo ($>6\times 10^{14} M_{\odot}$; \citealt{Giacintucci2017}) and the radio BCG coincides with the X-ray brightness peak, suggesting that this cluster may have harboured radio mini-halo in the past, but it is not possible to verify this possibility with the current observational data. We also note that, apart from the possible less merger energy injection in the off-axis region, the poor uv-coverage below 3 k$\lambda$ in the 1.7 GHz observation might also be responsible for the missing large scale halo emission. 
However, the procedure used for injections (Sect. \ref{subsec:spix_estimate}) already accounts for the interferometric response on scales that cover most of the halo emission at 325 MHz and allows to infer a limit to the spectral slope $\leq -1.52$. 

In comparison with other clusters hosting intermediate halos (e.g., \citealt{Bonafede2014,Sommer2017, Savini2018,Kale2019}, hereafter BSK), the SPT2031 is slightly different. The BSK clusters are all strong cool core clusters ($0.3 < c_{\mathrm{SB}} < 0.4$) whereas the SPT2031 has $c_{\mathrm{SB}} = 0.05$ indicating the absence of a cool core. However, the central cooling time of SPT2031, $t_{\mathrm{cool},0}=3.43$ Gyr suggests that it has a weak cool core. On the other hand, the centroid shift $w$ of all 4 clusters are $> 0.01$ which classifies them as disturbed and are potential host of radio halos. However, $w$ for SPT2031 lie around the disturbed/relaxed boundary and the photon asymmetry measurement ($A_{\mathrm{phot}}=0.25$) classifies it as a moderately disturbed cluster which is in agreement with the steep nature of the diffuse radio emission.

\section{Conclusions}
\label{sec:conclude}
We report the discovery of diffuse radio emission in the galaxy cluster SPT2031 with GMRT and VLA observations.
The size of this emission at 325 MHz was found to be about $\sim 600 - 800$ kpc.
The diffuse radio emission is present surrounding the central BCG, typical of radio mini-halos whereas the large size is comparable to those of the radio halos.
The dynamical state analysis reveals that it has a weak cool core along with disturbed morphology caused by a past merger event, i.e., it is in a transitional state between a merger and a cool core cluster. However, the current \textit{Chandra} X-ray data (10 ks) does not allow us to make a detailed study of the cluster dynamics for conclusive evidence.
The spectral index of the halo emission visible in both frequencies, mostly along the merger axis, was found to be $\alpha^{1679}_{325} = -1.35 \pm 0.07$ 
whereas steeper off-axis halo emission is largely undetected at 1.7 GHz.
We speculate the origin of this emission to be a past less energetic merger event in a cool core cluster where the observed intermediate radio halo is in transition into a future mini-halo.
Future sensitive radio observations at intermediate frequencies may shed light on the true spectral nature of the radio halo emission.
Additionally, deep X-ray observations along with resolved radio spectral index map are needed to study the ICM dynamics in detail for conclusive comment in favour of or against the above mentioned scenario.

\section*{Acknowledgements}
We would like to thank IIT Indore for providing necessary computing facilities for data analysis. We also thank both GMRT and VLA staff for making these radio observations possible. GMRT is run by the National Centre for Radio Astrophysics of the Tata Institute of Fundamental Research. RR is supported through ECR/2017/001296 grant awarded to AD by DST-SERB, India. MR would like to thank DST for INSPIRE fellowship program for financial support (IF160343). DR was supported by a NASA Postdoctoral Program Senior Fellowship at the NASA Ames Research Center, administered by the Universities Space Research Association under contract with NASA. This research was supported in part by NASA ADAP grant NNX15AE17G.











\bsp	
\label{lastpage}
\end{document}